\documentstyle[12pt,psfig]{article}
\psfigurepath{figs:figs/xfig:figs/con:figs/rig:figs/cayley}

\begin{document}
\

\leftline{\small Proceedings of \it Rigidity Theory and Applications}
\leftline{\small Traverse City, MI, June 14-18 1998}
\leftline{\small Fundamental Material Science Series, Plenum}
\vspace{4\baselineskip}

\noindent{\bf COMPARISON OF CONNECTIVITY AND RIGIDITY PERCOLATION}

\vspace{2 \baselineskip}

\newenvironment{add}
{\begin{list}{}{\setlength{\leftmargin}{1truein}}
      \item[]}
{\end{list}}
\begin{add}
Cristian F.~Moukarzel and Phillip M.~Duxbury\\
\\
Department of Physics/Astronomy and\\
Center for Fundamental Materials Research,\\
Michigan State University,\\
East Lansing, MI 48824-1116
\end{add}

\vspace{2\baselineskip}

\noindent{\bf 1. INTRODUCTION}
\vspace{\baselineskip}

Connectivity percolation has devotees in mathematics, physics and in myriad
applications~\hbox{[1-4]}.  Rigidity percolation is a more general problem,
with connectivity percolation as an important limiting case. There is a
growing realization that the general rigidity percolation problem exhibits a
broader range of fundamental phenomena and has the potential for many new
applications of percolation ideas and models. In connectivity percolation,
the propagation of a scalar quantity is monitored, while in rigidity the
propagation of a vector is, in general, considered. In both cases, one or
more conservation laws hold. Moreover, connectivity percolation is rather
special and appears to be, in many cases, quite different than other
problems in the general rigidity class.  We illustrate these differences by
comparing connectivity and rigidity percolation in two cases which are very
well understood, namely diluted triangular lattices~\hbox{[5,6]} and
trees~\hbox{[7]}.

A large part of the intense fundamental interest in percolation is due to the fact
that percolation is like a phase transition, in the sense that there is a
critical point (critical concentration) and non-trivial scaling behavior
near the critical point~\hbox{[1-4]}.  This analogy carries over to the
rigidity percolation problem.  However, although the connectivity
percolation problem is usually second order (including on trees), the
rigidity percolation problem is first order in mean field theory and on
trees~\hbox{[7]}.  However on triangular lattices, both connectivity and
rigidity percolation are second order, though they are in {\it different
universality classes}~\hbox{[5,6]}.  The emphasis of most of the analysis in
the literature and in this presentation is the behavior at and near the
critical point.

In this paper, we discuss (Section 2) ideas which apply to connectivity and
rigidity percolation on diluted lattices.  In Section 3, we discuss
and compare the specific case of connectivity and rigidity percolation on
trees.  In Section 4 we summarize the matching algorithms which may be used
to find the percolating cluster in both connectivity and rigidity cases. The
behavior in the connectivity and rigidity cases on site diluted triangular
lattices is then compared.  Section 5 contains a summary and discussion of
the similarities and differences between connectivity and rigidity percolation
in more general terms.

\noindent {\bf 2. RIGIDITY AND CONNECTIVITY OF RANDOM GRAPHS}\\

In mathematical terms,  percolation is the study of the connectivity
properties of random graphs as a
function of the number of edges in the graph.  We are usually interested in
the asymptotic limit of graphs with an infinite number of vertices.  In the
physics community, we usually study this process on a regular lattice (e.g.
square, simple cubic, triangular) and consider the effect of adding or
removing edges which lie between the vertices of these graphs.  However the
mean field limit is equivalent to considering connections between all sites
on the lattice, no matter their Euclidean separation.  The behavior of
this ``infinite range'' model is equivalent to the limit of infinite
dimensional lattice models (e.g. infinite dimensional hypercubes) and tree
models. The reason for the equivalence of the critical behavior of these
models is that they are all dominated by long range ``rings'', whereas on
regular lattices in lower dimensions short range rings can be very
important. Field theory models seek to add short range loops to these models
in a systematic and non-perturbative (in some sense) way.

We consider lattices consisting of sites which are defined to have $z$
neighbors (i.e. they have coordination number $z$). We then 
adding edges between neighbors randomly with probability $p$.  Once
this process is finished we are left with a random graph with each site
having average coordination $C=z p$.  In studying percolation, we are
always asking the question ``Is it possible to transmit some quantity (i.e.
a scalar, or a vector) across a graph''.  If it is
possible to transmit the quantity of interest, we say that the network is
above the ``percolation threshold'' $p_c$ (or equivalently $C_c=z p_c$) for
that quantity. If it is impossible to transmit the quantity of interest we
are below $p_c$. At $p_c$ the part of the random graph which transmits the
quantity of interest is {\it fractal} {\bf provided} the percolation
transition is second order. The probability that a bond is on this
``percolating backbone'' is $P_B$, and is one of the key quantities in the
analysis of percolation problems on regular lattices.  If the percolation
transition is ``second order'',  

\begin{equation} 
P_B \sim (p-p_c)^{\beta'} \ \ \hbox{as}\ \ p \to p_c^+ 
\end{equation} 

\noindent
with $\beta' > 0$.  On trees it is more difficult to define
$P_B$. Nevertheless we are able to analyze the problem effectively using
``constraint-counting'' ideas~\hbox{[7]}. 

Let us for a moment put aside thinking in terms of percolation and instead
develop constraint counting ideas originally discussed by Maxwell, and which
have been developed extensively in the engineering, math and glass
communities~\hbox{[7,8]}.  These ideas have not been applied to connectivity
percolation till recently and are enriching to both the connectivity and
rigidity cases.  At a conceptual level constraint counting is deceptively
simple.  To illustrate this, we assign to each site of our lattices a
certain number of {\it degrees of freedom}.  In connectivity percolation,
the transmission of a scalar quantity is of interest, therefore
each site is assigned one degree of freedom.  However
if we consider a point object in two dimensions from the point of view of
the transmission of forces, it has two degrees of freedom (two translations).
In $d$ dimensions point masses have $d$ degrees of freedom.  However extended
objects have rotational degrees of freedom, so that when we have clusters
which are mutually rigid, we must also consider these rotational degrees of
freedom. We call such objects bodies and they have $d(d-1)/2$ degrees of
freedom.  From a model viewpoint, we allow the number of degrees of freedom of
a free site to be a control variable which we label $g$, with $g=1$ the 
connectivity case. With equal generality we may say that $G$ is the number of
degrees of freedom of a rigid cluster or body. It is easy to see that there is
a vast array of models with $g\ne 1$, and with $G\ne 1$ 
and much of the physics and
applications of these models have yet to be examined~\hbox{[7]}. 
 
Now we have a model in which a free site has $g$ degrees of freedom, so that a
lattice with $N$ sites(and no edges) has a total of $F=Ng$ degrees of freedom,
or ``floppy modes'' (i.e. modes which have zero frequency due to the fact that
there is no restoring force).   
Constraint counting consists in simply saying
that each time an \emph{independent} edge is added to the lattice, the number
of degrees of freedom (zero frequency modes) is reduced by one if this edge
is not ``wasted'' (see later). For example, the
minimum number $E_{\hbox{\scriptsize \bf min}}$ of constraints needed to make a
rigid cluster out of a set of $N$ sites is

\begin{equation}
E_{\hbox{\scriptsize \bf min}} = N g - G,
\label{eq:bmin}
\end{equation}
which holds for the case in which edges are put on the graph in such
a way that none is wasted. In general, if $E$ edges are added to the graph,
the number of degrees of freedom (or floppy modes) which remains is  

\begin{equation}
F = N g - E + R. 
\end{equation}

Note the additional term $R$ on the right hand side.  This term is key in
understanding the relation between constraint counting and percolation, and
in finding algorithms for rigidity percolation. $R$ is the number of
``redundant bonds'' or ``wasted'' edges which are added to the lattice.  An
edge does not reduce the number of floppy modes if it is placed between two
sites which are already mutually rigid, in which case this edge is
``redundant''.  The simplest examples of subgraphs containing a redundant
bond on a triangular lattice are illustrated in Fig. 1 for the connectivity
($g=1$) and $g=2$ rigidity cases.

\begin{figure}[tbhp]
\centerline{ \psfig{figure=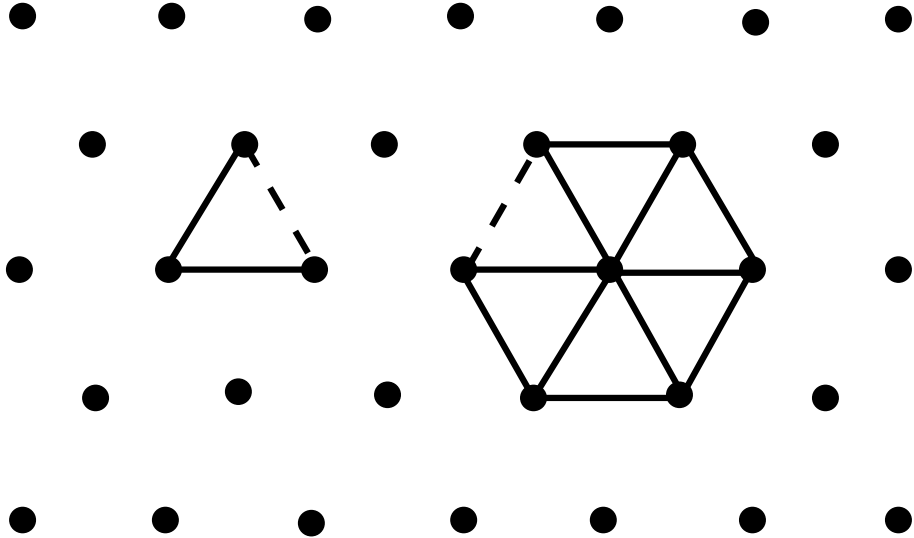,width=12cm,angle=270}}
\centerline{}
\caption{The simplest subgraphs on a triangular lattice which
contain a redundant bond (dashed).  Connectivity case (left), $g=2$ rigidity
case (right).}
\end{figure}

 Note that any one of the bonds in these structures could be labeled as the
redundant one.  However once any one of them is removed, all of the others
are necessary to ensure the mutual rigidity of the structure.  In fact the
set of all bonds which are mutually redundant form an ``overconstrained'' or
``stressed cluster''.  This is because, in the rigidity case we can think of
each edge as being a central force spring, which means that there is a
restoring force only in tension and compression.  Then an overconstrained
cluster of such springs (with random natural lengths) is internally
stressed due to the redundant bond. In the connectivity case each bond is
like a wire which can carry current or fluid flow. The simplest
overconstrained cluster is then a loop which can support an internal ``eddy''
current. Rigid structures which contain no redundant bonds are minimally rigid
or ``isostatic''. In connectivity percolation isostatic structures are trees,
whereas in ($g>1$) rigidity percolation isostatic structures must always
contain loops~(see Fig. 1).

In percolation problems, we are interested in the asymptotic limit of very
large graphs ($N \rightarrow \infty$), and it is more convenient to work
with intensive quantities, so we define $f(p) = F/gN$ and $r(p)=2Rg/zN$
which leads to

\begin{equation}
f(p) = 1 - {z\over 2g}( p  - r(p)),
\end{equation}
where the number of edges $E/N = zp/2$.  The normalization on $r(p)$ is
chosen this way because $r(p)$ is now the probability that a bond is
redundant~(times $g$).  Note that we normalize the number of floppy modes by
$g$ to be consistent with previous work~\hbox{[5-7]}. We have shown that
$f(p)$ acts as a free energy for both connectivity and rigidity problems, so
that if $\partial f(p)/\partial p$ undergoes a jump then the
transition is first order.  The behavior of this quantity is directly
related to the probability that a bond is overconstrained $P_{ov}$ via the
important relation~\hbox{[6]} 

\begin{equation} 
{\partial f \over \partial p}
= -{z\over 2g} ( 1 - P_{ov}) 
\end{equation}

If the transition is second order, the second derivative $\partial^2
f/\partial p^2 \sim (p-p_c)^{-\alpha}$, where $\alpha$ is the specific heat
exponent~\hbox{[6]}.

On both triangular lattices and on trees, we also calculate the infinite
cluster probability.  This is composed of the backbone plus
the dangling ends. Dangling ends are rigidly
connected to the backbone but {\it do not} participate in the transmission
of the quantity of interest. Examples of dangling ends in the connectivity and
rigidity cases are illustrated in Fig. 2.

\begin{figure}[tbhp]
\centerline{ \psfig{figure=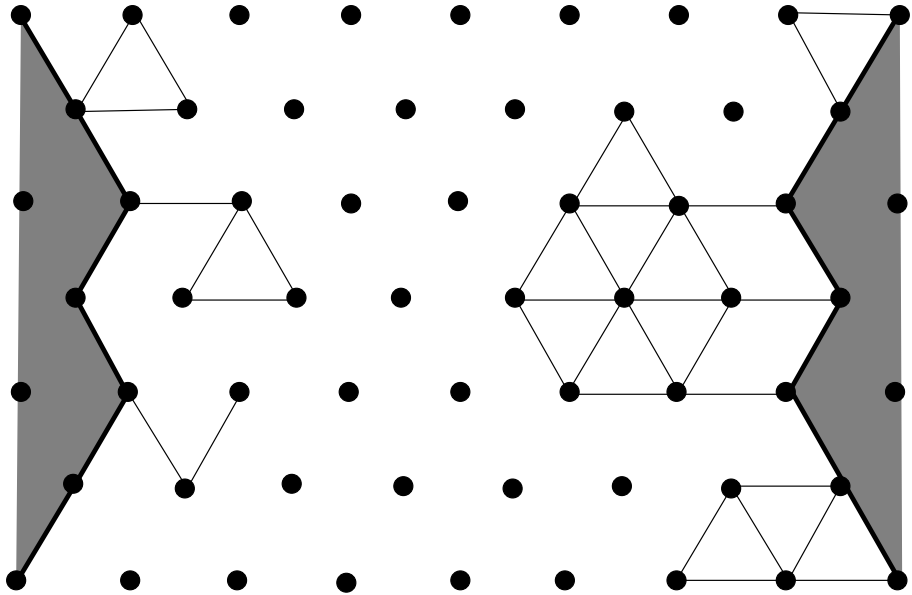,width=12cm,angle=270}}
\centerline{}
\caption{ 
Examples of dangling ends (thin lines) connected to a backbone
(shaded). Connectivity case (left) and $g=2$ rigidity case (right).
}
\end{figure}

\noindent {\bf 3. TREES}\\

In our previous analysis of tree models, we have considered propagation of
rigidity outward from a rigid boundary (this is the same as a strong surface
field). Here we also add a bulk field which induces rigidity.  The results
are essentially the same whether the boundary or the bulk field is used. In
percolation problems, the bulk field corresponds to nuclei embedded in the
lattice and tethered to a rigid substrate.  This is the sort of construction
envisioned by Obukov~\hbox{[9]}, and used extensively in the construction of
models for connectivity percolation.  The mean-field equation for the order
parameter on trees is then,

\begin{equation}
T_0= h + (1-h) \sum^{z-1}_{l=g} {z-1 \choose l}
(pT_0)^l (1-pT_0)^{z-1 -l},
\label{eq:mf}
\end{equation}
where $T_0$ is the probability that a site is connected to the infinite
rigid cluster through one branch of a tree (see the paper by Leath
for more details on the derivation), and $h$ is the 
probability that a site is rigidly connected to the 
rigid substrate ($h$ is sometimes called a ``ghost field'').
On trees, the probability that a bond is overconstrained is,

\begin{equation}
{N_0\over N_B} = T_0^2
\end{equation}
so that we have the key equation,

\begin{equation}
{\partial f \over \partial p} = -{z\over 2g} (1-T_0^2).
\end{equation}

Other useful formulae are the probability that a bond
is overconstrained,

\begin{equation}
P_{ov} = T_0^2,
\end{equation}
and the probability that a bond is on the infinite cluster,

\begin{equation}
P_{inf} = T_0^2 + 2T_0 T_1.
\end{equation}

In the connectivity cases, an infinitesimal field $h$ or any finite order
parameter at the boundary is sufficient to allow a percolation transition to
occur on trees.  In contrast in $g\ne 1$ cases, there must be a finite $h$,
or a finite boundary field before a transition occurs.  These differences
are illustrated in Fig. 3.

\begin{figure}[tbhp]
\vskip -1cm
\hbox{\hsize=0.5\hsize
\centerline{ \psfig{figure=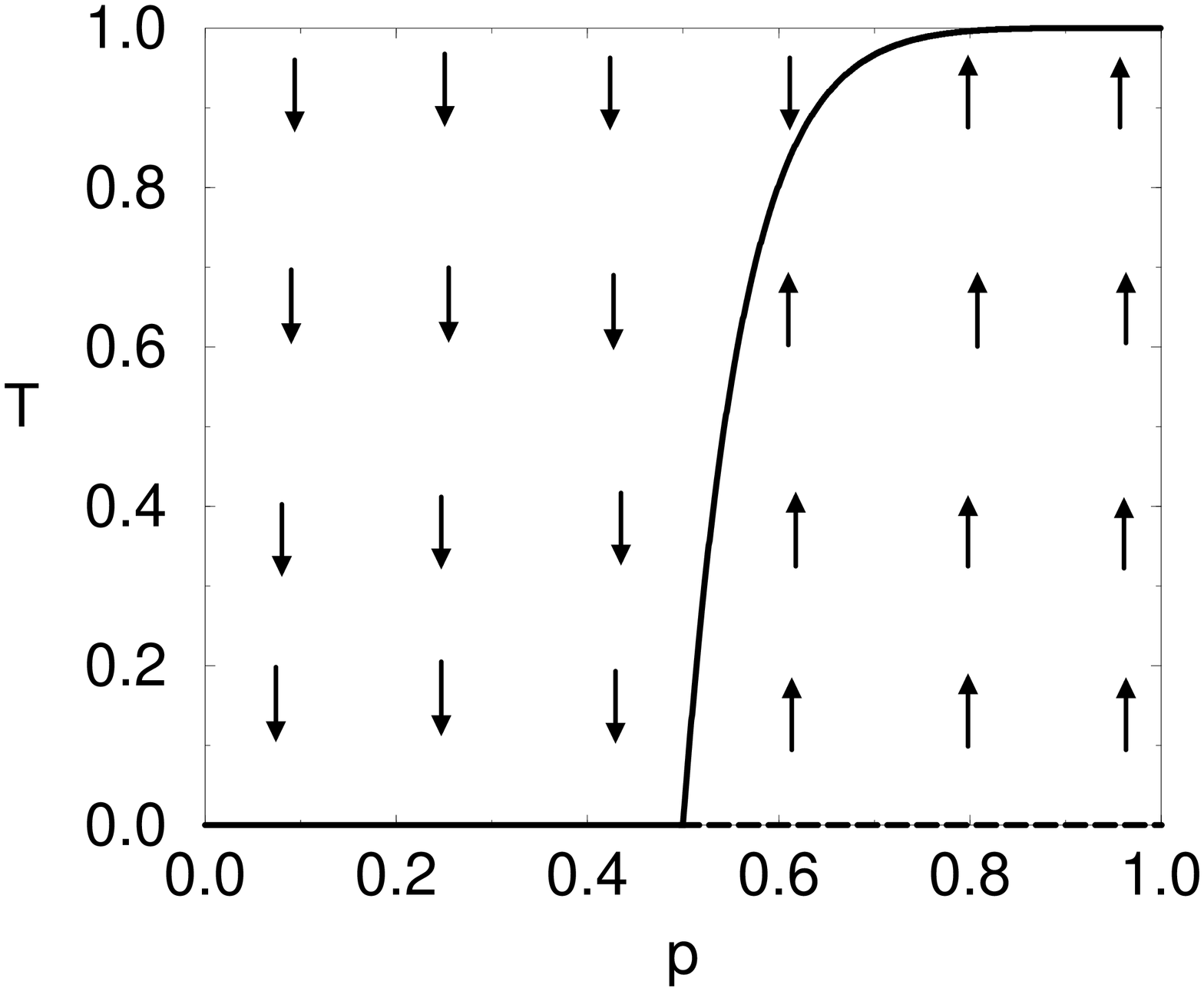,width=8cm,angle=270}}
\centerline{ \psfig{figure=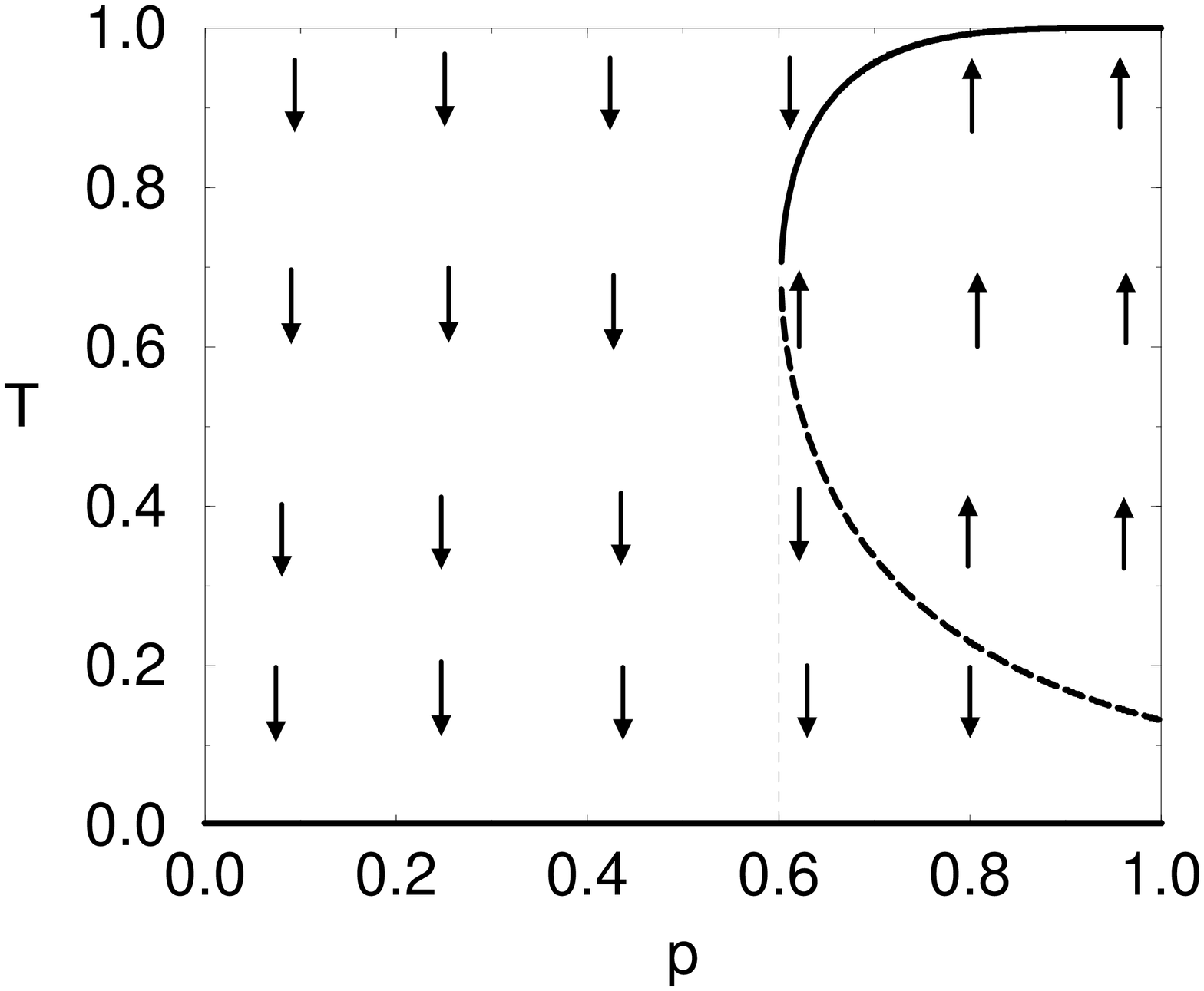,width=8cm,angle=270}}
}
\vskip 1cm
\caption{Domains of attraction of the mean field equations (\ref{eq:mf}) with
$h=0$ : (left) A typical connectivity behavior (this example is $z=3$); and (right)
a typical rigidity behavior (this example is $z=6$, $g=2$). Dark lines are 
stable (attractive) solutions and shaded lines repulsive. 
}
\end{figure}

It is clearly seen from these figures that the rigidity transition is first
order on trees, while the connectivity one is second order. The results
presented in Fig. 3 are simple to obtain.  We iterate the mean field
equation (6) until a stable fixed point is reached (there are similar
equations for $T_1$ etc.) and we then evaluate Eqs. (7) - (9).  We identify
the point at which the stable solution becomes nonzero as $p_s$, the spinodal
point, for reasons discussed below.  In the connectivity case $p_s = p_c$
because the transition is second order.

We also want to find the total number of redundant bonds $r(p)$ and the
total number of floppy modes $f(p)$.  In order to find these quantities, we
integrate Eq. (7) and then use Eq. (3).  However, the integration of Eq. (7)
leads to one free constant.  This constant depends on the situation we wish
to model.  If we wish to model a regular lattice, then we impose the
constraint,

\begin{equation}
r(1) = 1 - {2g \over b z},
\label{eq:weuse}
\end{equation}
for example in the case of central force springs on a triangular lattice
$1/3$ of the springs are redundant when $p=1$. However on trees,

\begin{equation}
r(1) = 1 - {g \over b (z-1)}.
\end{equation}

In using tree models to provide approximations to rigidity percolation on
regular lattices, we impose the constraint (\ref{eq:weuse})~\hbox{[7]}.  
Then we find
that r(p) approaches zero at a critical point $p_c$, which is NOT the same
a $p_s$ if $g>1$.  However, if $g>1$, this $p_c$ is very close to the
Maxwell estimate $p_c^{\hbox{\tiny Maxwell}} = 2g/bz$, and is {\it the same}
as that found numerically for infinite range (random bond) models~\hbox{[7]}.
Typical results found for tree models are presented in Fig. 4.

\begin{figure}[tbhp]
\vskip -0.7cm
\hbox{\hsize=0.5\hsize
\centerline{ \psfig{figure=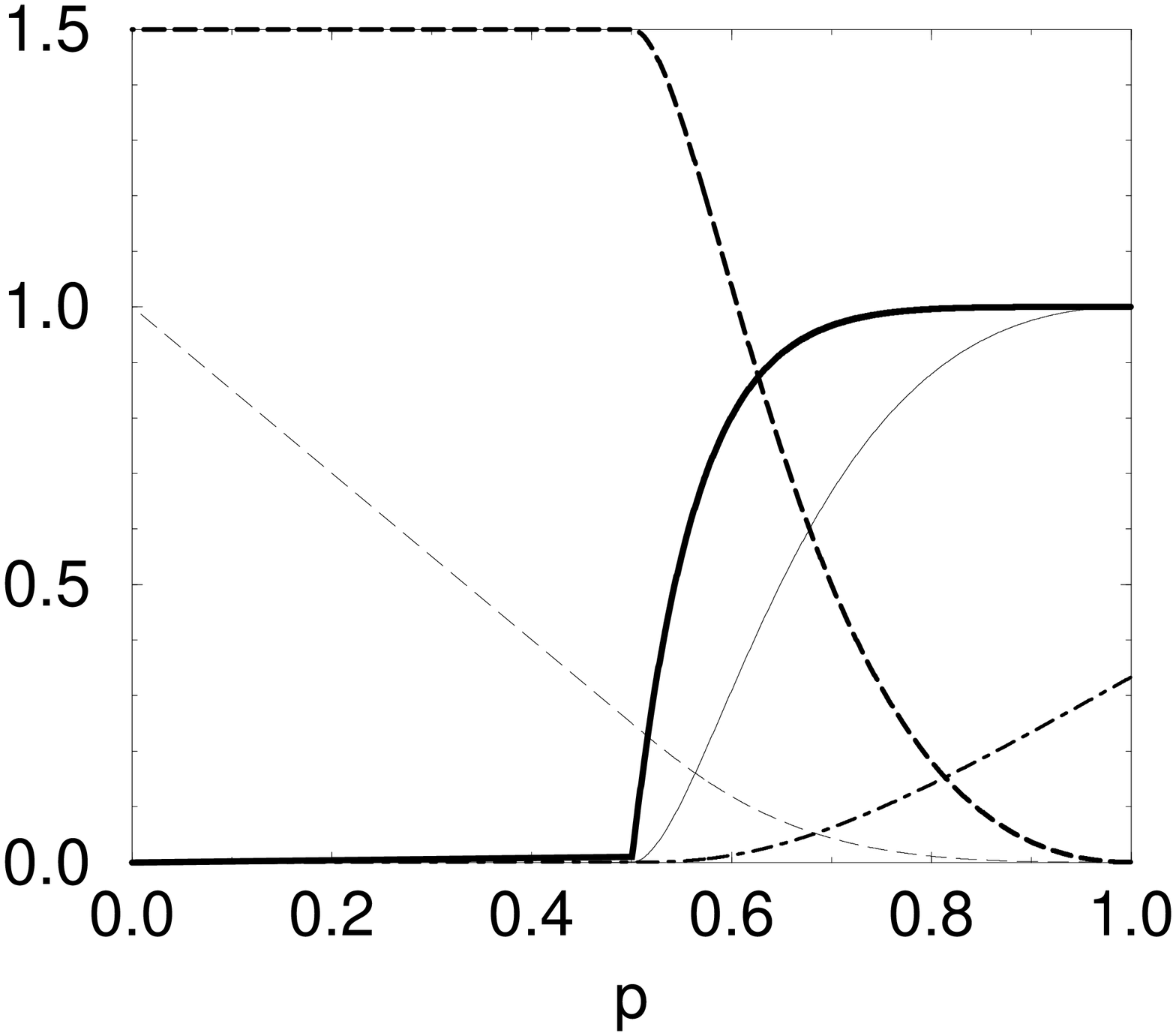,width=8cm,angle=270}}
\centerline{ \psfig{figure=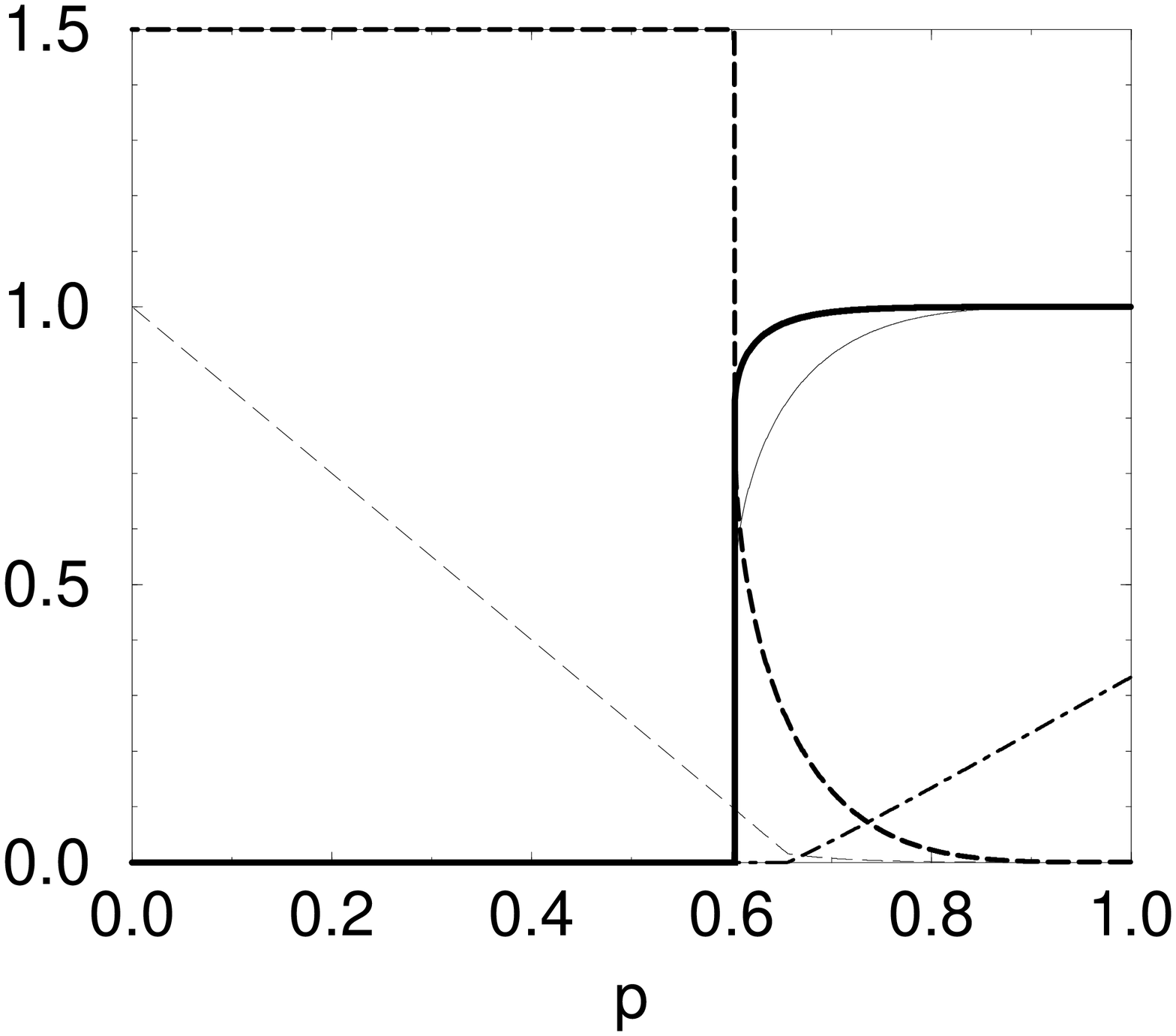,width=8cm,angle=270}}
}
\vskip 0.7cm
\caption{The order parameter, floppy modes and its derivative
as a function of $p$ for a typical connectivity $z=3,g=1$(left),
and rigidity $z=6,g=2$(right) cases, both with $b=1$. $f(p)$(dashed),
$r(p)$(dot dashed), $-df/dp$(heavy dashed), $P_{ov}$(thin solid),
$P_{inf}$(heavy solid)}
\end{figure}

In the connectivity case we find $p_c=p_s$, while in rigidity cases
$p_s<p_c$.  For the rigidity case of Fig. 4, $p_s=0.605$, while
$p_c=0.655$.\\

\noindent {\bf 4. TRIANGULAR LATTICES}\\

We consider connectivity percolation($g=1,b=1,z=6$) and central force
percolation($g=2,b=1,z=6$) on site diluted triangular lattices.  In
connectivity percolation the scaling behavior near the percolation threshold
has been proven on trees and has been extensively tested on regular lattices
using large scale numerical simulations.  We have carried out a similar
program for rigidity percolation as summarized below for triangular
lattices.  However doing numerical simulations in the rigidity case has
required a breakthrough in algorithm development, and this has only occurred
recently through contact with the mathematical computer science community. 
The methods developed for rigidity have even improved some aspects of
algorithmic methods~\hbox{[12]} for the connectivity case, as discussed below.

\begin{figure}[tbhp]
\vbox{
\hbox{\hsize=0.5\hsize
\centerline{ \psfig{figure=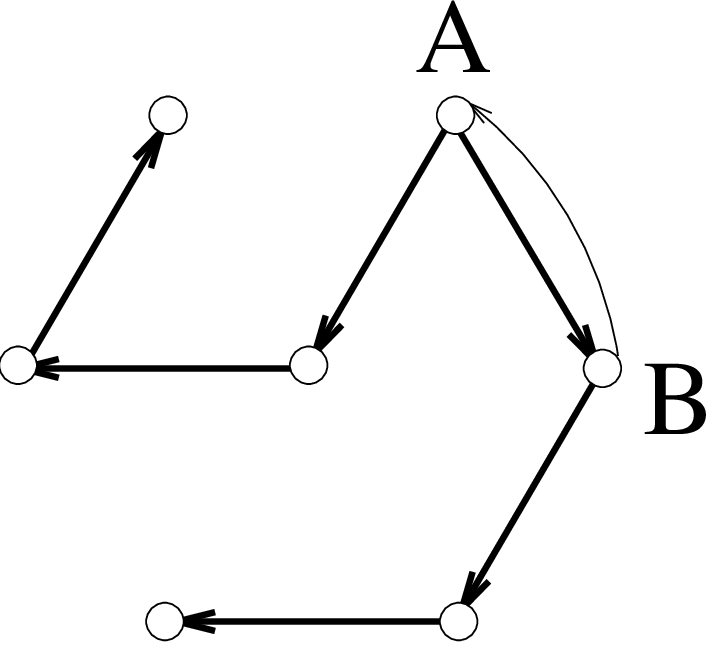,width=5cm,angle=270}}
\centerline{ \psfig{figure=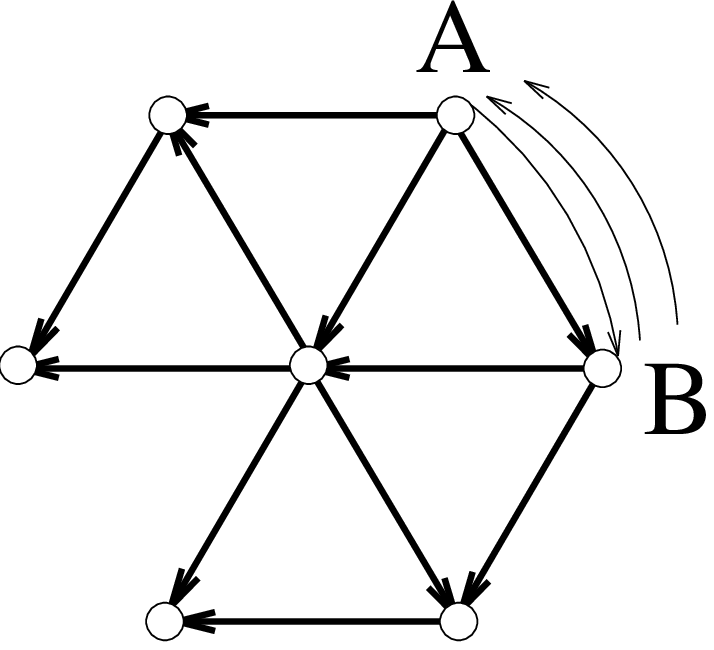,width=5cm,angle=270}}
}
\vspace{1cm}
\hbox{\hsize=0.5\hsize
\centerline{ \psfig{figure=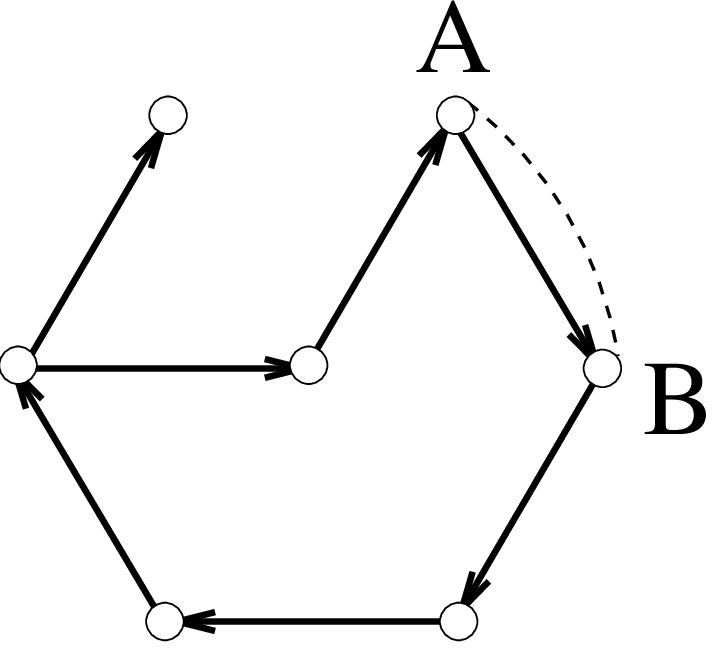,width=5cm,angle=270}}
\centerline{ \psfig{figure=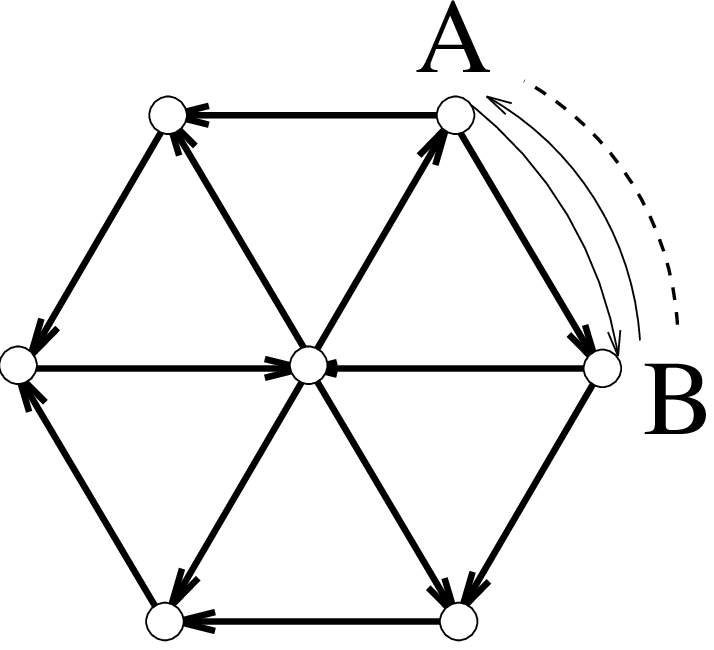,width=5cm,angle=270}}
}
}
\vspace{0.3cm}
\caption{Examples of successful (top figures) and
failed (bottom figures) matches in the connectivity
( $g=G=1$, left figures) and joint-bar rigidity ( $g=2,  G=3$, right figures) cases 
on triangular lattices. AB is the new bond that is being tested.
Each site has $g$ degrees of freedom and therefore accepts at most $g$ incoming
arrows. Each new bond carries with it $G$ auxiliary arrows, which must
also be matched. If any of the arrows cannot be matched (dashed), the new bond
is redundant. }
\end{figure}

Before discussing the rigidity algorithms, it is important to point out the
limitations of these methods.  Firstly these algorithms are able to identify 
structures which can support stress (rigidity case), or which
can carry a current (connectivity case).  They do not find the actual current
or stress, but rather those bonds which are able to transmit the load from an
input node or set of nodes to an output node or set of nodes.  In the
connectivity percolation case it is trivial that the actual geometric
realization of a given graph connectivity does not change the set of bonds
which carry current from a fixed set of input nodes to a fixed set of output
nodes. However in the rigidity case, there can be ``special'' realizations
of a given graph which are responsible for singular rigid properties of
``generically'' rigid clusters. 
This may occur when there are ``degenerate''
constraints (e.g. parallel bonds on a lattice).  The probability of such
degenerate realizations is zero on geometrically disordered lattices and for
this reason geometry, and the existence of these degenerate configurations, 
may be simply ignored in this case. In the mathematical
community this problem is called ``generic rigidity'' and is the only one
for which powerful algorithms exist.  Thus the problem of rigidity on a {\it
regular triangular lattice remains unsolved}.  The results described below
apply to triangular lattices whose sites have been randomly displaced (e.g.
by $0.1$ of a lattice constant).  In addition, the powerful matching
algorithms that we use apply to ``joint-bar graphs in the plane'' and to a
subset of graphs in general dimensions 
(so-called body - bar problems) However, it turns out that glasses in $3-d$ 
correspond to a case which is solvable (they map to a body-bar problem), so
this is one of those unusual cases where the practical case is actually
theoretically convenient.
\begin{figure}[tbhp]
\hbox{\hsize=0.5\hsize
\centerline{ \psfig{figure=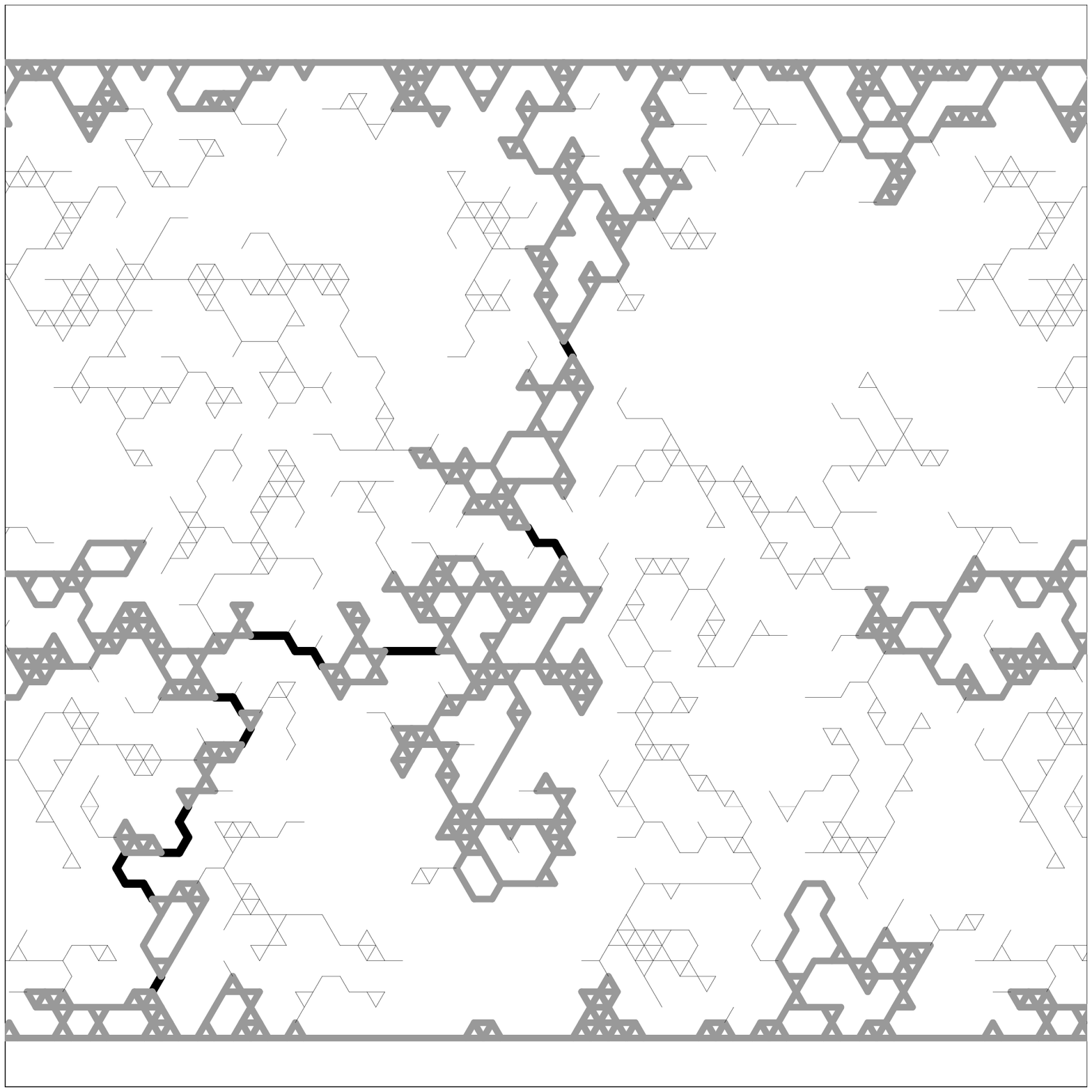,width=8cm}}
\centerline{ \psfig{figure=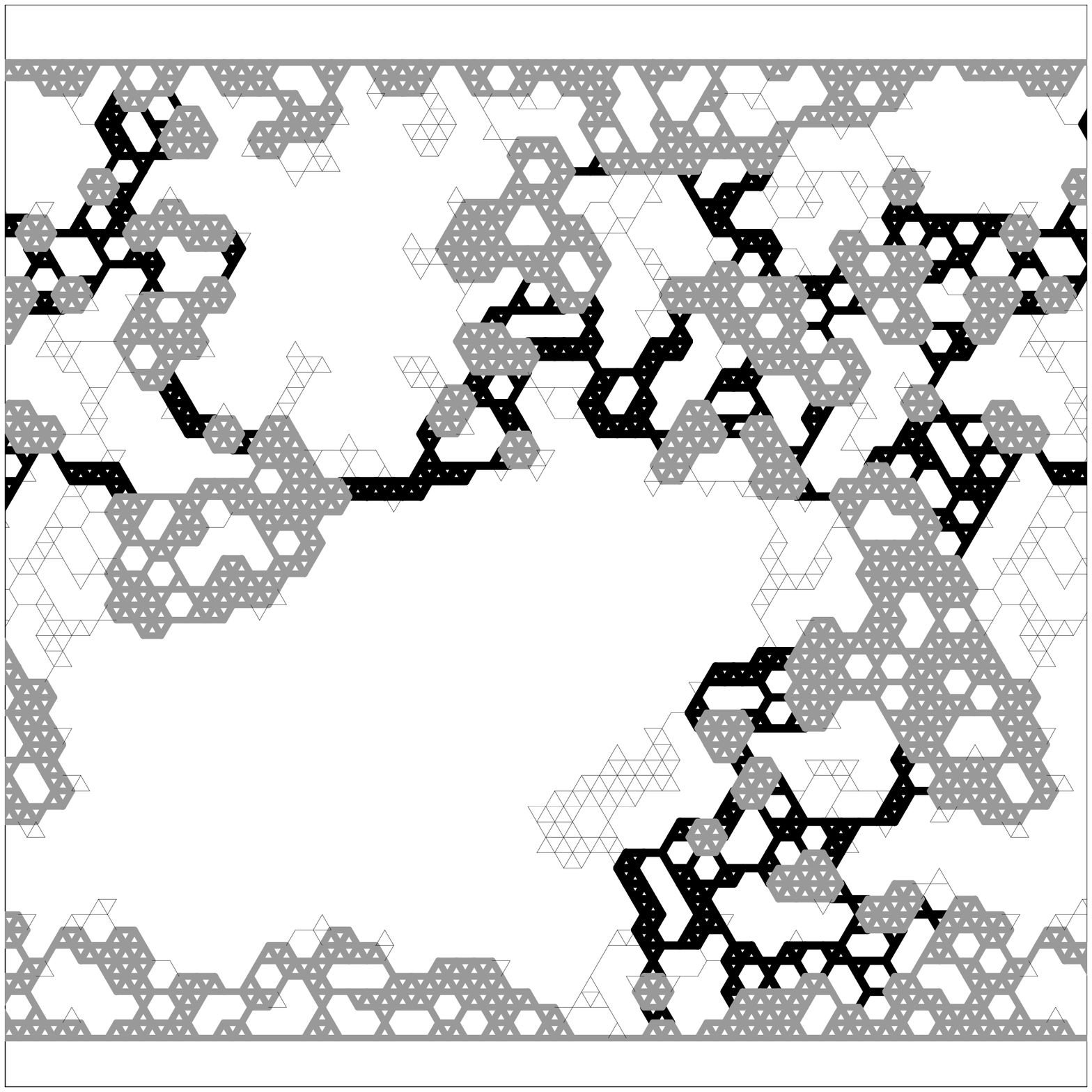,width=8cm}}
}
\caption{The infinite-cluster geometry for connectivity
(left) and rigidity(right) percolation
on site-diluted triangular lattices with $L=64$.  In each case dark wide
lines are cutting bonds, wide lines are non-critical backbone bonds (blobs)
and thin lines are dangling ends.  }
\end{figure}

\noindent
The matching algorithm is implemented as follows~\hbox{[10]}:

\begin{quote}
Start with an empty triangular lattice(no bonds) and
assign to each node $g$ degrees of freedom. 
\\ \noindent
{\it Then:}
\begin{enumerate}
\item Randomly add a bond to the lattice.
\item Test whether this bond is redundant with respect
to the bonds which are currently in the lattice.
\item If the bond is  redundant {\it do not add it
to the lattice}, but instead store its location in
a different array.
\item Return to 1.
\end{enumerate}
{\it End}
\end{quote}

The key step is step 2.  The algorithm to do this test is rigorous and based
on Laman's theorem. It was developed by Bruce Hendrickson, who also provided a
key service in explaining his algorithm to the physics community. Step 2. is
performed by exact constraint counting.  
This is implemented by ``matching'' constraints (bonds) to degrees of freedom, with
the restriction that the constraints can only match degrees of freedom at
each end of the corresponding bond.  Thus it is natural to represent this constraint
counting by using arrows to indicate the degree of freedom to which each
bond is matched.  The idea is that, when a new bond is added, it must be
possible to match $G+1$ arrows to degrees of freedom of the graph in order for this new bond
not to be redundant. If this task can be accomplished~\footnote{It is valid
to rearrange previously existing arrows, provided this is done in such a way
that all remain matched to some degree of freedom}
the new bond is {\it independent}, which means that
it is not wasted, and is left on the graph. In this case only $G$ auxiliary
arrows are removed. If on the other hand some of the new arrows cannot be
matched, the last edge is {\it redundant} and all $G+1$ arrows are erased.

A successful and a failed match are illustrated in Fig. 5 for a
connectivity (g=1) case and a rigidity (g=2) case.  
Note that the bond that is being tested carries with it $G$ additional
``copies'' which account for global degrees of freedom of a rigid cluster.  In the
connectivity case $G=1$, while on central-force bar-joint networks in two
dimensions $G=3$.

A failed match occurs when a bond is unable to find a degree of freedom to
``cover''.  This bond is then redundant.  In trying to find a degree of freedom
to which a redundant bond can be assigned or ``matched'', the algorithm
identifies all bonds which are ``overconstrained'' of stressed with respect
to that bond.  This set of bonds is called a {\it Laman subgraph}.  Note that
if a redundant bond is already in a graph, it is not possible to add a new
bond and test its redundancy with this method.  This is the reason that the
algorithm proceeds by adding bonds one at a time starting with an empty
lattice.  Any error is testing the redundancy of a bond invalidates the rest
of the addition sequence.  However since this algorithm is an integer method
there is no problem with roundoff.  It is easy to see that the matching
algorithm is quite efficient, however if requires quite a bit of effort to
fully optimize these methods.

Pictures of the infinite cluster for connectivity and rigidity percolation
on a triangular lattice are presented in Fig. 6.
\begin{figure}[tbhp]
\vskip -1cm
\hbox{\hsize=0.5\hsize
\centerline{ \psfig{figure=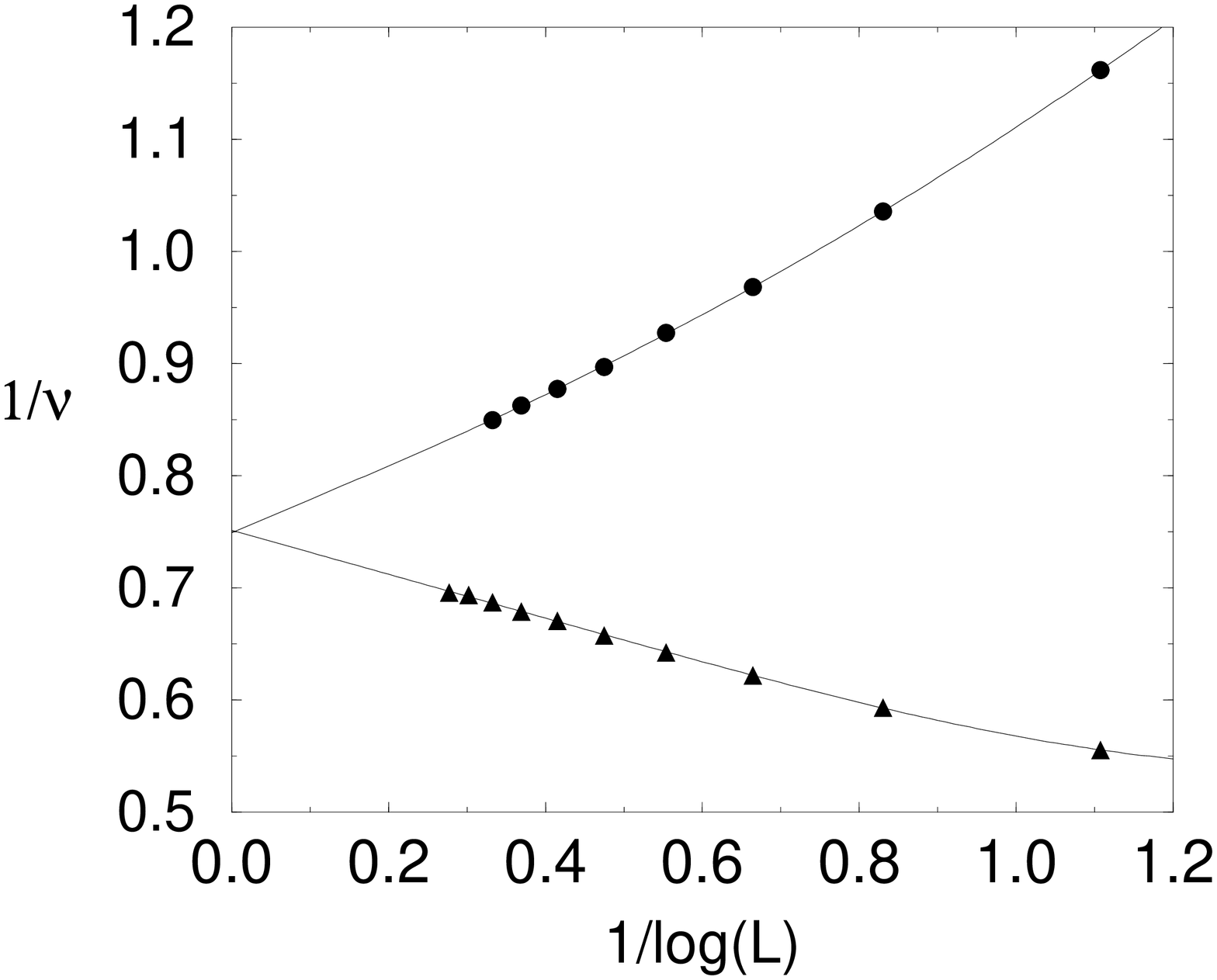,width=8cm,angle=270}}
\centerline{ \psfig{figure=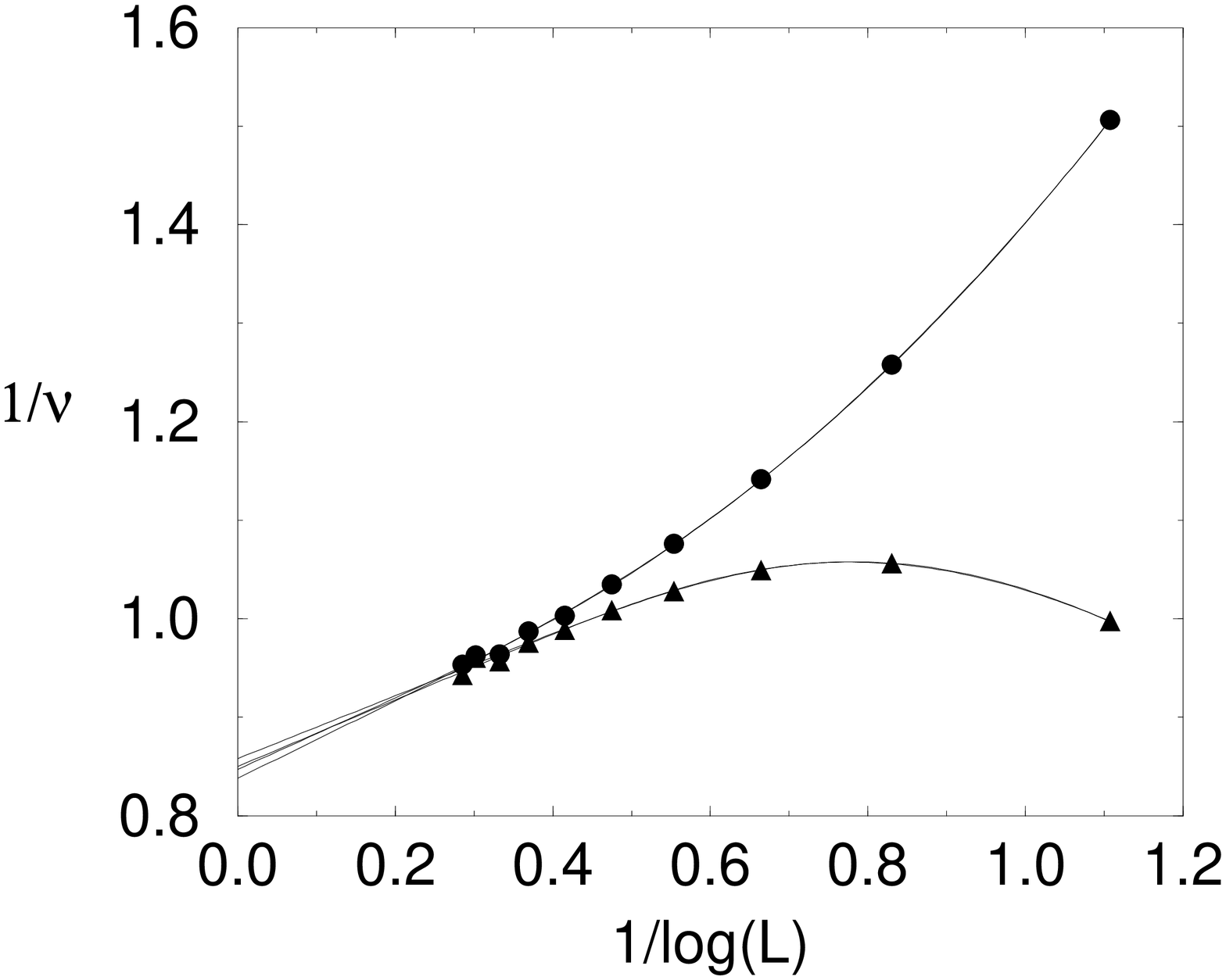,width=8cm,angle=270}}
}
\vskip 1cm
\caption{
Analysis to find the correlation length exponent $\nu$
for connectivity (left figure) and rigidity percolation (right
figure) on triangular lattices.  In both figures the upper curve is a fit
(including corrections to scaling) of  $\delta p_c \sim L^{-1/\nu}$ (full
circles) while the  lower  curve is for the number
of cutting bonds $n_c \sim L^{1/\nu}$ (full triangles). } 
\end{figure}

Because of the fact that  we add bonds one at a time until the percolation
point is reached, we are able to identify $p_c$ {\it exactly} for
each sample, and therefore measure the components of the spanning
cluster exactly at $p_c$. This eliminates the error associated with
measurements at fixed values of $p$, since estimated exponents are known
to depend very sensitively on $p$. This method was proposed and applied for
the first time in ref.~\hbox{[5]}.

At $p_c$ we identify three different types of bonds: {\it backbone bonds}, {\it
dangling ends} and {\it cutting bonds}.  These together form the {\it infinite
cluster}. The cutting bonds are stressed
(belong to the backbone), but they are ``critical'' because if one of them is
removed, load is no longer 
transmitted across the infinite cluster.  The results in Fig. 5 are for bus
bars at the top and bottom of the figures and periodic boundary conditions
in the other direction. In order to find the correlation length exponent, we
used two relations: Firstly the size dependence of the threshold behaves as
$\delta p_c \sim L^{-1/\nu}$ and secondly, the number of cutting bonds
varies as $n_c \sim L^{1/\nu}$~\hbox{[4]}.  An analysis of this data in the
connectivity and rigidity cases is presented in Fig. 7.

It is seen from this figure that the rigidity case has a different
$\nu=1.16\pm 0.03$~\hbox{[5,6]} than the connectivity case $\nu=4/3$.  A
finite-size-scaling plot of the infinite-cluster and backbone densities at
$p_c$ is presented in Fig. 8, along with the density of dangling ends.

\begin{figure}[tbhp]
\vskip -0.7cm
\hbox{\hsize=0.5\hsize
\centerline{ \psfig{figure=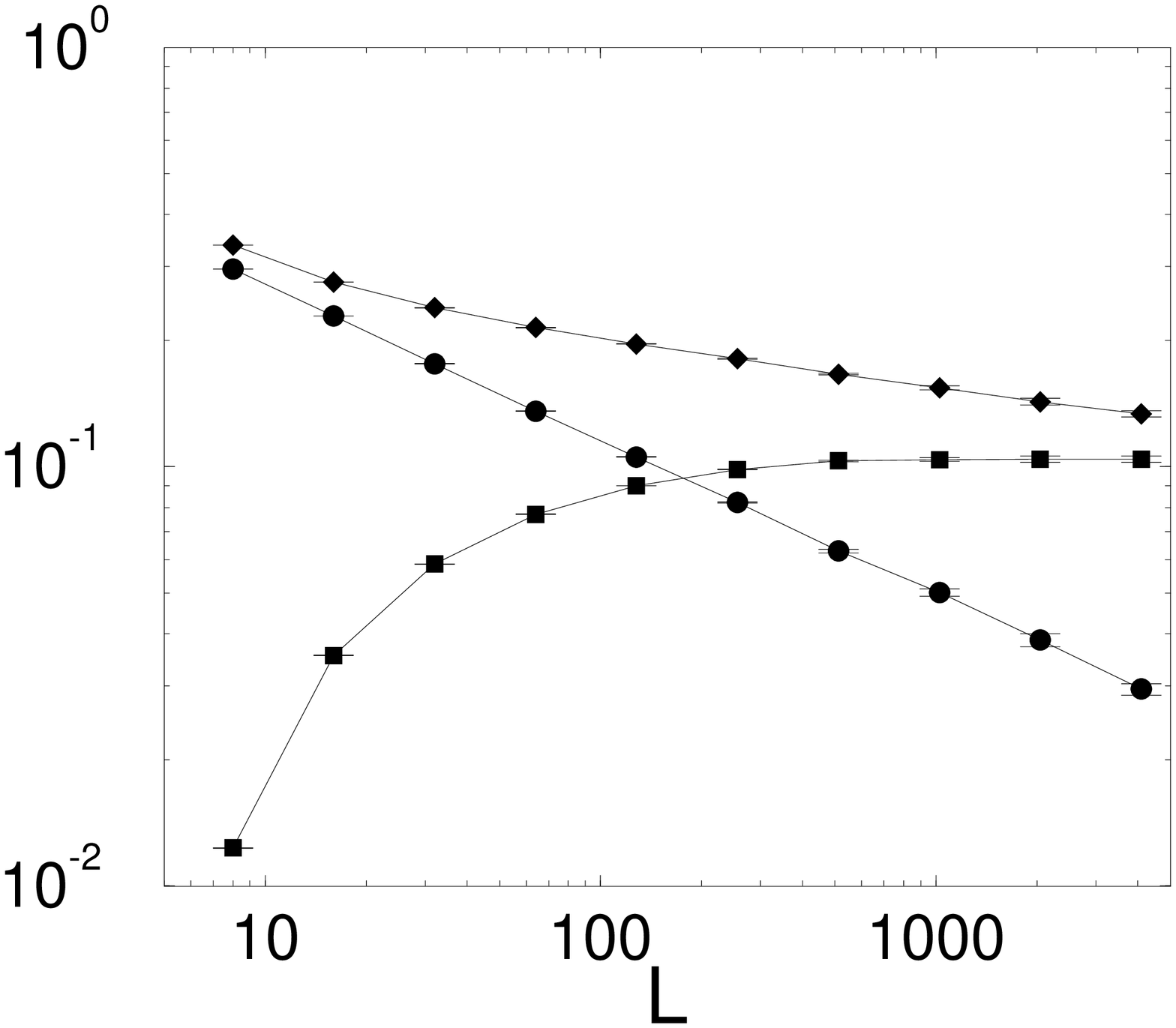,width=8cm,angle=270}}
\centerline{ \psfig{figure=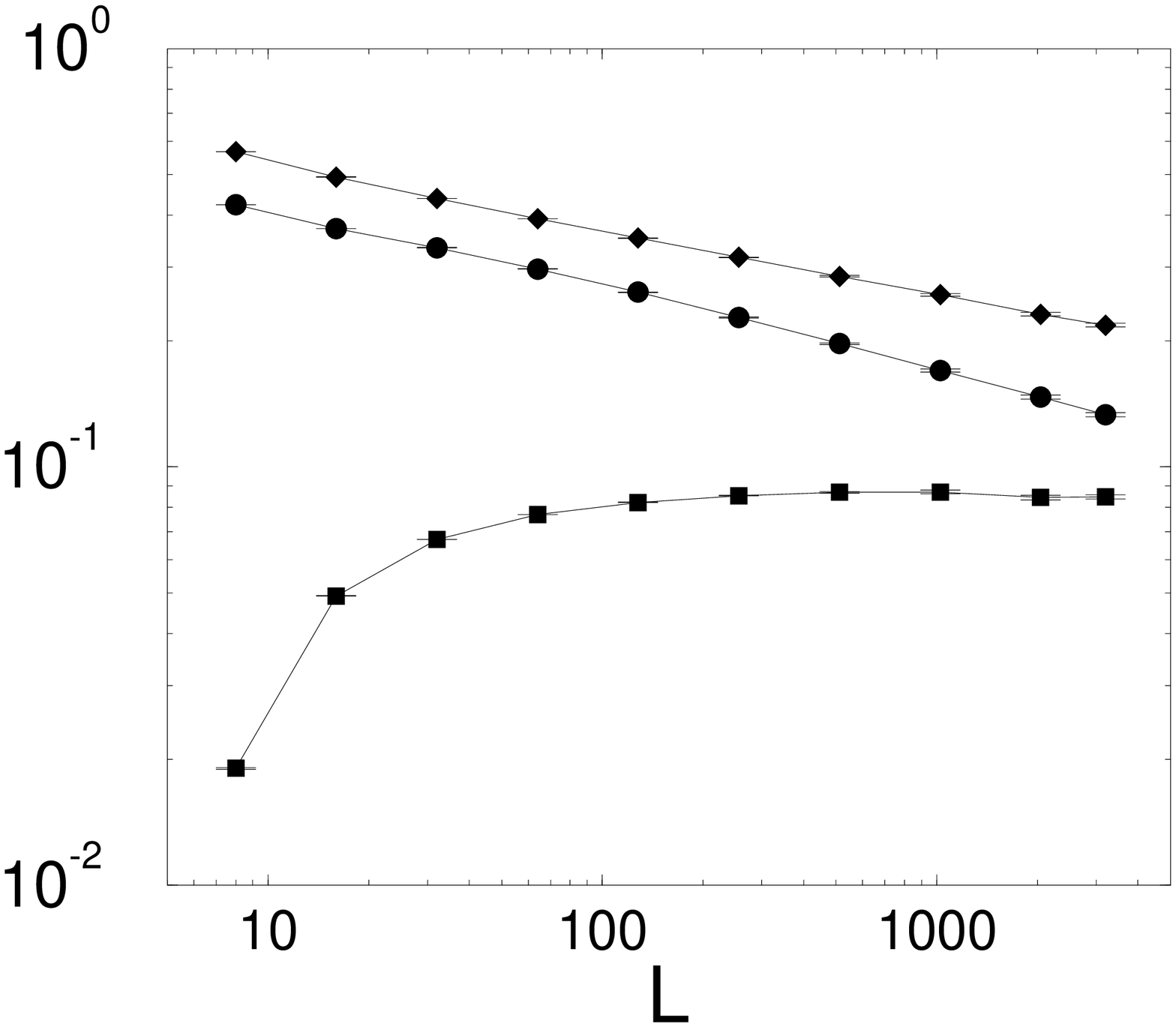,width=8cm,angle=270}}
}
\vskip 0.7cm
\caption{
The density of backbone bonds(circles), 
infinite cluster bonds(diamonds) and dangling bonds(squares) 
for the connectivity ( $g=1$, left figure)
and rigidity ( $g=2$, right figure) cases on site diluted lattices.
Our method of single-bond addition allowed these measurements
to be made {\it exactly} at $p_c$ for each sample.
}
\end{figure}

What is unambiguous from these figures is that the backbone density is
decreasing algebraically $P_B \sim L^{-\beta'/\nu}$, and from this data we
find that in the rigidity case $\beta' = 0.25 \pm 0.02$~\hbox{[5,6]}.  It
also appears that the infinite-cluster probability is decreasing
algebraically, however that is difficult to reconcile with the behavior of
the dangling ends.  In the connectivity case much larger scale simulations
have been done, and it is known that the algebraic decrease in the infinite
cluster probability continues, so that the infinite cluster in the
connectivity case is indeed fractal~\hbox{[4]}. In the rigidity case, such
large scale simulations are still lacking, so the possibility of an infinite
cluster with finite density (about $0.1$) remains.\\

\noindent {\bf 5. SUMMARY}\\

The development of tree models and the use of new algorithms in numerical
studies have revolutionized our understanding of the geometry of rigidity
percolation. In particular we now know that on triangular lattices, the
rigidity transition is second order, but in a different universality class
to the connectivity case. In contrast on trees the rigidity transition is
first order while the connectivity transition is second order. The geometry
of rigidity percolation is different from that of connectivity percolation
due to the requirement of multiple connectivity in the rigidity case. 
However this is clearly not enough to ensure that the transition becomes
first order.  Perhaps a deeper question is whether the infinite cluster
breaks up into two or an infinite number of subclusters when a critical bond
is removed.  In the connectivity case, the answer is clearly two.  In the
rigidity case it is infinity on trees and difficult to analyze precisely on
triangular lattices.  This and a host of other questions remain unanswered in
this interesting class of problems.\\

\noindent {\bf ACKNOWLEDGEMENTS}

PMD and CM thank the DOE for support under
contract DE-FG02-90ER45418 and CM also acknowledges support 
from the Conselho Nacional de Pesquisa CNPq, Brazil.

\vspace{2\baselineskip}

\vbox{
\noindent{\bf REFERENCES}\\

\list
  {\relax}{\setlength{\labelsep}{0em}
        \setlength{\itemindent}{-\parindent}
        \setlength{\leftmargin}{\parindent}}
    \def\newblock{\hskip .11em plus .33em minus .07em}
    \sloppy\clubpenalty4000\widowpenalty4000
    \sfcode`\.=1000\relax%

\small
\setlength{\parskip}{-4pt}

{
\item{} 1. \ 
G.~Deutscher, R.~Zallen and J.~Adler (Eds.), 
{\it Percolation Structures and Processes}, Bristol: Adam Hilger (1983)
\item{} 2. \ 
G.~Grimmett, {\it Percolation}, New York: Springer-Verlag (1989)
\item{} 3. \ 
H.~Kesten, {\it Percolation Theory for Mathematicians},
Boston:  Birkhäuser (1982)
\item{} 4. \ 
D.~Stauffer and A.~Aharony,  {\it Introduction to Percolation Theory}
2nd ed., London: Taylor \& Francis (1992)
\item{} 5. \
C.~Moukarzel and P.M.~Duxbury, Phys.~Rev.~Lett. {\bf 75},
4055 (1995);  
C.~Moukarzel, P.M.~Duxbury and P.L.~Leath, 
Phys.~Rev.~Lett. {\bf 78}, 1480 (1997) 
\item {} 6. \ 
D.~Jacobs and M.F.~Thorpe, Phys.~Rev.~Lett. {\bf 75}, 4051 (1995); 
D.~Jacobs and M.F.~Thorpe, Phys.~Rev. {\bf E53}, 3682 (1996)
\item{} 7. \
C.~Moukarzel, P.M.~Duxbury and P.L.~Leath, Phys.~Rev. {\bf E55}, 5800 
(1997); P.M.~Duxbury, D.~Jacobs, M.F.~Thorpe and C.~Moukarzel,
submitted to Phys.~Rev.~E.
\item{} 8. \
J.C.~Phillips, J.~Non-Cryst.~Solids {\bf 43}, 37 (1981);
M.F.~Thorpe, J.~Non-Cryst.~Sol. {\bf 57}, 355 (1983)
\item{} 9. \
S.~Obukov, Phys.~Rev.~Lett. {\bf 74}, 4472 (1994)
\item{} 10. \
B.~Hendrickson, Siam J.~Comput. {\bf 21},65 (1992);
C.~Moukarzel, J.~Phys. {\bf A29}, 8079 (1996)
\item{} 11. \
Comment to PRL by D.~Jacobs and M.F.~Thorpe,  
and reply by P.M.~Duxbury, C.~Moukarzel and P.L.~Leath.} to appear in the 
June 15th issue of Phys.~Rev.~Lett. 
\item{} 12. \
C.~Moukarzel 1997, Int.~J.~Mod.~Phys.~C, to appear.
}
\end{document}